\documentclass[%
 aip,
 amsmath,amssymb,
 reprint,%
]{revtex4-1}

\usepackage{graphicx}
\usepackage{dcolumn}
\usepackage{bm}

\usepackage[utf8]{inputenc}
\usepackage[T1]{fontenc}
\usepackage{mathptmx}
\usepackage{etoolbox}

\usepackage{natbib} 
\usepackage[usenames,dvipsnames]{color}
\usepackage{amsmath} 
\usepackage{amssymb} 
\usepackage{soul} 
\usepackage{ifthen} 
\usepackage[ampersand]{easylist} 
\usepackage{todonotes}
\usepackage{hyperref}

\def\be#1\ee{\begin{equation}#1\end{equation}}
\def\ba#1\ea{\begin{align}#1\end{align}}
\def\bg#1\eg{\begin{gather}#1\end{gather}}

\def\shownote{1} 
\newcommand{\nt}[1]{\ifthenelse{\shownote=1}{\textcolor{Red}{[[#1]]}}{}}

\makeatletter
\def\@email#1#2{%
 \endgroup
 \patchcmd{\titleblock@produce}
  {\frontmatter@RRAPformat}
  {\frontmatter@RRAPformat{\produce@RRAP{*#1\href{mailto:#2}{#2}}}\frontmatter@RRAPformat}
  {}{}
}%
\makeatother
\begin{document}

\preprint{AIP/123-QED}

\title[Investigation of coherence of niobium-based resonators enabled by a fast-sealing microwave cavity]{Investigation of coherence of niobium-based resonators enabled by a fast-sealing microwave cavity} 
\author{C. Zhang}
\author{R. Germond}%
\author{N. Janzen}
\affiliation{Institute for Quantum Computing, University of Waterloo, Waterloo, ON N2L 3G1, Canada}%
\affiliation{Department of Physics and Astronomy, University of Waterloo, ON N2L 3G1, Canada}

\author{A.-M. Valente-Feliciano}
\affiliation{Jefferson Lab, 12000 Jefferson Ave, Newport News, VA 23606, USA.}

\author{M. Bal}
\affiliation{Superconducting Quantum Materials and Systems Division, Fermi National Accelerator Laboratory (FNAL), Batavia, IL 60510, USA.}%

\author{A. Lupascu}
  \email{c76zhang@uwaterloo.ca, adrian.lupascu@uwaterloo.ca}
\affiliation{Institute for Quantum Computing, University of Waterloo, Waterloo, ON N2L 3G1, Canada}%
\affiliation{Department of Physics and Astronomy, University of Waterloo, ON N2L 3G1, Canada}
\affiliation{Waterloo Institute for Nanotechnology, University of Waterloo, Waterloo, ON, N2L 3G1, Canada}

\date{\today}

\begin{abstract}
Resonators and qubits with a niobium (Nb) base metal layer achieve some of the highest coherence times in superconducting quantum devices. The performance of such devices is often limited by loss associated with two-level systems, which are found primarily at material surfaces and interfaces. The metal-air (MA) interface is a major contributor to device loss. In this work, we develop a fast-sealing microwave cavity that enables devices to be placed under vacuum within five minutes of oxide removal, thereby significantly reducing the MA interface loss compared to common device processing and packaging approaches. Using coplanar stripline resonators, we demonstrate that devices sealed in such a cavity exhibit internal quality factors exceeding one million at single-photon power. After re-exposure to air, the devices show downward resonance frequency shifts and quality factor degradations, quantitatively consistent with a model of Nb oxide regrowth. The fast-sealing microwave cavity provides a practical and consistent method to mitigate MA interface loss and sustain high coherence in Nb devices, and establishes a controlled platform for studying metal oxide regrowth kinetics and dielectric properties, the understanding of which is critical to achieving high coherence in superconducting quantum devices.
\end{abstract}

\maketitle

Superconducting quantum circuits are among the most promising platforms for realizing fault-tolerant quantum computing\cite{Kjaergaard_Schwartz_Braumuller_Krantz_Wang_Gustavsson_Oliver_2020}. Since their inception, device coherence times have improved by close to six orders of magnitude\cite{Kjaergaard_Schwartz_Braumuller_Krantz_Wang_Gustavsson_Oliver_2020, Jiang_Deng_Fan_Li_Sun_Tan_Wang_Xue_Yan_Yu_et_al._2025}, with state-of-the-art qubits approaching millisecond coherence times\cite{Place_Rodgers_Mundada_Smitham_Fitzpatrick_Leng_Premkumar_Bryon_Vrajitoarea_Sussman_et_al._2021a, Wang_Li_Xu_Li_Wang_Yang_Mi_Liang_Su_Yang_et_al._2022b, Deng_Song_Gao_Xia_Bao_Jiang_Ku_Li_Ma_Qin_et_al._2023, Somoroff_Ficheux_Mencia_Xiong_Kuzmin_Manucharyan_2023, Bal_Murthy_Zhu_Crisa_You_Huang_Roy_Lee_Zanten_Pilipenko_et_al._2024, Biznarova_Osman_Rehnman_Chayanun_Krizan_Malmberg_Rommel_Warren_Delsing_Yurgens_et_al._2024, Tuokkola_Sunada_Kivijarvi_Albanese_Gronberg_Kaikkonen_Vesterinen_Govenius_Mottonen_2025, Bland_Bahrami_Martinez_Prestegaard_Smitham_Joshi_Hedrick_Pakpour-Tabrizi_Kumar_Jindal_et_al._2025} and resonator quality factors exceeding one million\cite{Verjauw_Potocnik_Mongillo_Acharya_Mohiyaddin_Simion_Pacco_Ivanov_Wan_Vanleenhove_et_al._2021, Altoe_Banerjee_Berk_Hajr_Schwartzberg_Song_Alghadeer_Aloni_Elowson_Kreikebaum_et_al._2022,  Crowley_McLellan_Dutta_Shumiya_Place_Le_Gang_Madhavan_Bland_Chang_et_al._2023, Lozano_Mongillo_Piao_Couet_Wan_Canvel_Vadiraj_Ivanov_Verjauw_Acharya_et_al._2024, Dhundhwal_Duan_Brauch_Arabi_Fuchs_Haghighirad_Welle_Scharwaechter_Pal_Scheffler_et_al._2025, Marcaud_Perello_Chen_Umbarkar_Weiland_Gao_Diez_Ly_Mahuli_D_Souza_et_al._2025, Bruckmoser_Koch_Tsitsilin_Grammer_Bunch_Richard_Schirk_Wallner_Feigl_Schneider_et_al._2025}. A key avenue for further coherence improvement lies in the choice of base-metal material and the optimization of associated fabrication processes\cite{De_Leon_Itoh_Kim_Mehta_Northup_Paik_Palmer_Samarth_Sangtawesin_Steuerman_2021, Murray_2021, Siddiqi_2021}. The highest-coherence devices to date employ aluminum\cite{Somoroff_Ficheux_Mencia_Xiong_Kuzmin_Manucharyan_2023, Biznarova_Osman_Rehnman_Chayanun_Krizan_Malmberg_Rommel_Warren_Delsing_Yurgens_et_al._2024}, tantalum\cite{Place_Rodgers_Mundada_Smitham_Fitzpatrick_Leng_Premkumar_Bryon_Vrajitoarea_Sussman_et_al._2021a, Wang_Li_Xu_Li_Wang_Yang_Mi_Liang_Su_Yang_et_al._2022b, Crowley_McLellan_Dutta_Shumiya_Place_Le_Gang_Madhavan_Bland_Chang_et_al._2023, Bland_Bahrami_Martinez_Prestegaard_Smitham_Joshi_Hedrick_Pakpour-Tabrizi_Kumar_Jindal_et_al._2025}, niobium (Nb)\cite{Verjauw_Potocnik_Mongillo_Acharya_Mohiyaddin_Simion_Pacco_Ivanov_Wan_Vanleenhove_et_al._2021, Altoe_Banerjee_Berk_Hajr_Schwartzberg_Song_Alghadeer_Aloni_Elowson_Kreikebaum_et_al._2022, Bal_Murthy_Zhu_Crisa_You_Huang_Roy_Lee_Zanten_Pilipenko_et_al._2024, Tuokkola_Sunada_Kivijarvi_Albanese_Gronberg_Kaikkonen_Vesterinen_Govenius_Mottonen_2025, Bruckmoser_Koch_Tsitsilin_Grammer_Bunch_Richard_Schirk_Wallner_Feigl_Schneider_et_al._2025}, or titanium nitride\cite{Deng_Song_Gao_Xia_Bao_Jiang_Ku_Li_Ma_Qin_et_al._2023} as the base superconducting metal layer.

Among the above materials, Nb offers several unique advantages. Nb has a high critical temperature\cite{Kittel_2005} and low kinetic inductance\cite{Annunziata_Santavicca_Frunzio_Catelani_Rooks_Frydman_Prober_2010}, leading to a low thermal quasiparticle density at millikelvin temperatures\cite{Kaplan_Chi_Langenberg_Chang_Jafarey_Scalapino_1976}, enhanced stability against temperature fluctuations\cite{De_Ory_Rollano_Rodriguez_Magaz_Granados_Gomez_2025}, and low device-to-device variability\cite{Verjauw_Potocnik_Mongillo_Acharya_Mohiyaddin_Simion_Pacco_Ivanov_Wan_Vanleenhove_et_al._2021}. Nb fabrication is also compatible with industrial-scale processing of circuits\cite{Tolpygo_Bolkhovsky_Weir_Johnson_Gouker_Oliver_2015}, and has synergy with Nb-based Josephson junctions\cite{Anferov_Lee_Zhao_Simon_Schuster_2024}. These advantages have driven significant interest in Nb-based high-coherence quantum devices.

However, Nb-based device performance is often limited by loss associated with two-level systems (TLSs)\cite{Muller_Cole_Lisenfeld_2019, McRae_Wang_Gao_Vissers_Brecht_Dunsworth_Pappas_Mutus_2020}. These TLSs mostly originate in amorphous surfaces and interfaces of the thin film\cite{Gao_Daal_Vayonakis_Kumar_Zmuidzinas_Sadoulet_Mazin_Day_Leduc_2008, Kumar_Gao_Zmuidzinas_Mazin_LeDuc_Day_2008, Burnett_Faoro_Lindstrom_2016, Niepce_Burnett_Latorre_Bylander_2020, Verjauw_Potocnik_Mongillo_Acharya_Mohiyaddin_Simion_Pacco_Ivanov_Wan_Vanleenhove_et_al._2021, Altoe_Banerjee_Berk_Hajr_Schwartzberg_Song_Alghadeer_Aloni_Elowson_Kreikebaum_et_al._2022,  Bal_Murthy_Zhu_Crisa_You_Huang_Roy_Lee_Zanten_Pilipenko_et_al._2024}, with especially large contributions from the lossy native oxides\cite{Burnett_Faoro_Lindstrom_2016,  Verjauw_Potocnik_Mongillo_Acharya_Mohiyaddin_Simion_Pacco_Ivanov_Wan_Vanleenhove_et_al._2021, Altoe_Banerjee_Berk_Hajr_Schwartzberg_Song_Alghadeer_Aloni_Elowson_Kreikebaum_et_al._2022, Bal_Murthy_Zhu_Crisa_You_Huang_Roy_Lee_Zanten_Pilipenko_et_al._2024}. Strategies to reduce oxide-related loss include etching with hydrofluoric (HF) acid\cite{Verjauw_Potocnik_Mongillo_Acharya_Mohiyaddin_Simion_Pacco_Ivanov_Wan_Vanleenhove_et_al._2021,Altoe_Banerjee_Berk_Hajr_Schwartzberg_Song_Alghadeer_Aloni_Elowson_Kreikebaum_et_al._2022}, vacuum baking\cite{Romanenko_Pilipenko_Zorzetti_Frolov_Awida_Belomestnykh_Posen_Grassellino_2020, Bafia_Murthy_Grassellino_Romanenko_2024}, surface encapsulation\cite{De_Ory_Rollano_Rodriguez_Magaz_Granados_Gomez_2025, Bal_Murthy_Zhu_Crisa_You_Huang_Roy_Lee_Zanten_Pilipenko_et_al._2024}, and post-deposition N\(_2\) passivation\cite{Zheng_Kowsari_Thobaben_Du_Song_Ran_Henriksen_Wisbey_Murch_2022, Fang_Oh_Kramer_Romanenko_Grassellino_Zasadzinski_Zhou_2023}. While oxide removal by HF acid etching or vacuum baking can yield close to an order-of-magnitude improvement in device quality factor\cite{Verjauw_Potocnik_Mongillo_Acharya_Mohiyaddin_Simion_Pacco_Ivanov_Wan_Vanleenhove_et_al._2021, Romanenko_Pilipenko_Zorzetti_Frolov_Awida_Belomestnykh_Posen_Grassellino_2020}, these gains are short-lived due to the rapid regrowth of the Nb oxides\cite{Verjauw_Potocnik_Mongillo_Acharya_Mohiyaddin_Simion_Pacco_Ivanov_Wan_Vanleenhove_et_al._2021, Grundner_Halbritter_1984}. Encapsulation circumvents this problem but leaves the metal sidewalls, where electric fields are the strongest, unprotected\cite{Chang_Shumiya_McLellan_Zhang_Bland_Bahrami_Mun_Zhou_Kisslinger_Cheng_et_al._2025}. Devices passivated by N\(_2\) experience quality factor degradation after long-time storage\cite{Zheng_Kowsari_Thobaben_Du_Song_Ran_Henriksen_Wisbey_Murch_2022}. Finding methods to consistently preserve the sample metal-air (MA) interface after surface treatment therefore remains an important and active pursuit.

In this letter, we develop a fast-sealing microwave cavity that can be evacuated to sub-millibar pressures within 5 minutes of sample etching by HF acid, thereby limiting oxide regrowth. To demonstrate the cavity's effectiveness, we fabricate Nb coplanar stripline (CPS) resonators with various geometries on a sapphire substrate, and test device quality factors using our approach of oxide removal and packaging. For comparison, we measure the quality factor of the resonators evacuated in the cavity immediately after etching, and re-measure the same resonators after breaking vacuum and allowing the sample to re-oxidize under ambient conditions for 8 days. Before re-exposure, the highest-coherence device exhibits a quality factor of \((2.5\pm0.3)\times10^6\) at single-photon power. After re-exposure, we observe device resonance frequency shifts and quality factor degradation, both showing a systematic trend with device geometry. Analysis of these quantities allows us to extract the regrown Nb oxide's thickness and loss tangent.

The fast-sealing microwave cavity is formed by two complementary copper mating parts with a combined internal volume of \(30\times24.5\times5.5\) mm\(^3\) (Fig. 1(a)). Copper is chosen as the cavity material because it enables the operation of flux-tunable devices in the future and does not induce a significant contribution to resonator loss. The sample sits in a pocket on the cavity bottom part's mating surface, and is held in place by small indium blobs placed in the pocket underneath the sample. The cavity's seam is sealed with an indium wire mounted in a groove in the bottom cavity part. A knife-edge on the top part aids in compressing the indium. A pinch-off tube, soldered to the cavity bottom part, connects to the cavity internal volume via a small access hole and allows for pumping and leak checking. An SMA connector is mounted through a hole in the cavity top part and is hermetically sealed with Stycast 2850FT.

For device packaging, each part of the cavity is first fully assembled, with indium preloaded in the bottom part. After HF acid etching and the ensuing water baths, the sample is blow-dried and immediately loaded and sealed in the cavity in the cleanroom. The cavity is then connected via a tube-to-KF-16 adapter to a nearby pumping station and evacuated. Across three separate runs, the time from removing the sample from the water bath to evacuating the cavity is under 5 minutes in each run. The evacuated cavity is then isolated with a closed valve and transported for leak checking (Fig. 1(c)). Leak rates have been measured between \(10^{-10}\) and \(10^{-8}\) mbar\(\cdot\)L/s. The cavity leak rate for the devices reported here is measured at \(1.0\times 10^{-10}\) mbar\(\cdot\)L/s. Assuming air leaking into the cavity at this rate, Nb\(_2\)O\(_5\) growth kinetics from literature\cite{Verjauw_Potocnik_Mongillo_Acharya_Mohiyaddin_Simion_Pacco_Ivanov_Wan_Vanleenhove_et_al._2021} imply \(<1\) pm oxide regrowth for over a month. The pinch-off tube is finally crimped to form a seal. The leak rate from the crimped end of the pinch-off tube has been tested by itself to be on the \(10^{-10}\) mbar\(\cdot\)L/s level. The fully sealed cavity is then installed in the dilution refrigerator while sample oxide regrowth is being suppressed.

The cavity is designed for microwave measurements in reflection mode. The reflection is given by
\begin{equation}
    S_{11}(\omega)=Ce^{i\left(\phi_0+\omega t_\mathrm{ed}\right)}\frac{2i(\omega-\omega_0)-\omega_0/Q_\mathrm{ext}+\omega_0/Q_\mathrm{int}}{2i(\omega-\omega_0)+\omega_0/Q_\mathrm{ext}+\omega_0/Q_\mathrm{int}} \, ,\label{eq:S11}
\end{equation}
where \(\omega\) is the probe frequency, \(\omega_0=2\pi f_0\) is the mode resonance frequency, \(C\) is the magnitude of the reflection baseline, \(\phi_0\) is the phase offset, \(t_\mathrm{ed}\) is the electrical delay, and \(Q_\mathrm{int}\) and \(Q_\mathrm{ext}\) are the internal and external quality factors\cite{Nguyen_2020}. Aided by ANSYS HFSS\cite{ansys} simulations, the location and center-pin length of the SMA connector are carefully chosen such that \(Q_\mathrm{int}\) and \(Q_\mathrm{ext}\) are comparable to each other for both the cavity mode and device modes\footnote{Simulation details in the supplementary information.\label{supplementary_information}}. Room-temperature testing of the cavity fundamental mode shows \(f_0=7.687\) GHz, \(Q_\mathrm{int}=4700\), and \(Q_\mathrm{ext}=7100\) (Fig. 1(b)), agreeing well with the simulated values of \(f_0=7.605\) GHz, \(Q_\mathrm{int}=5100\), and \(Q_\mathrm{ext}=6300\).

\begin{figure}
    \centering
    \includegraphics[width=\linewidth]{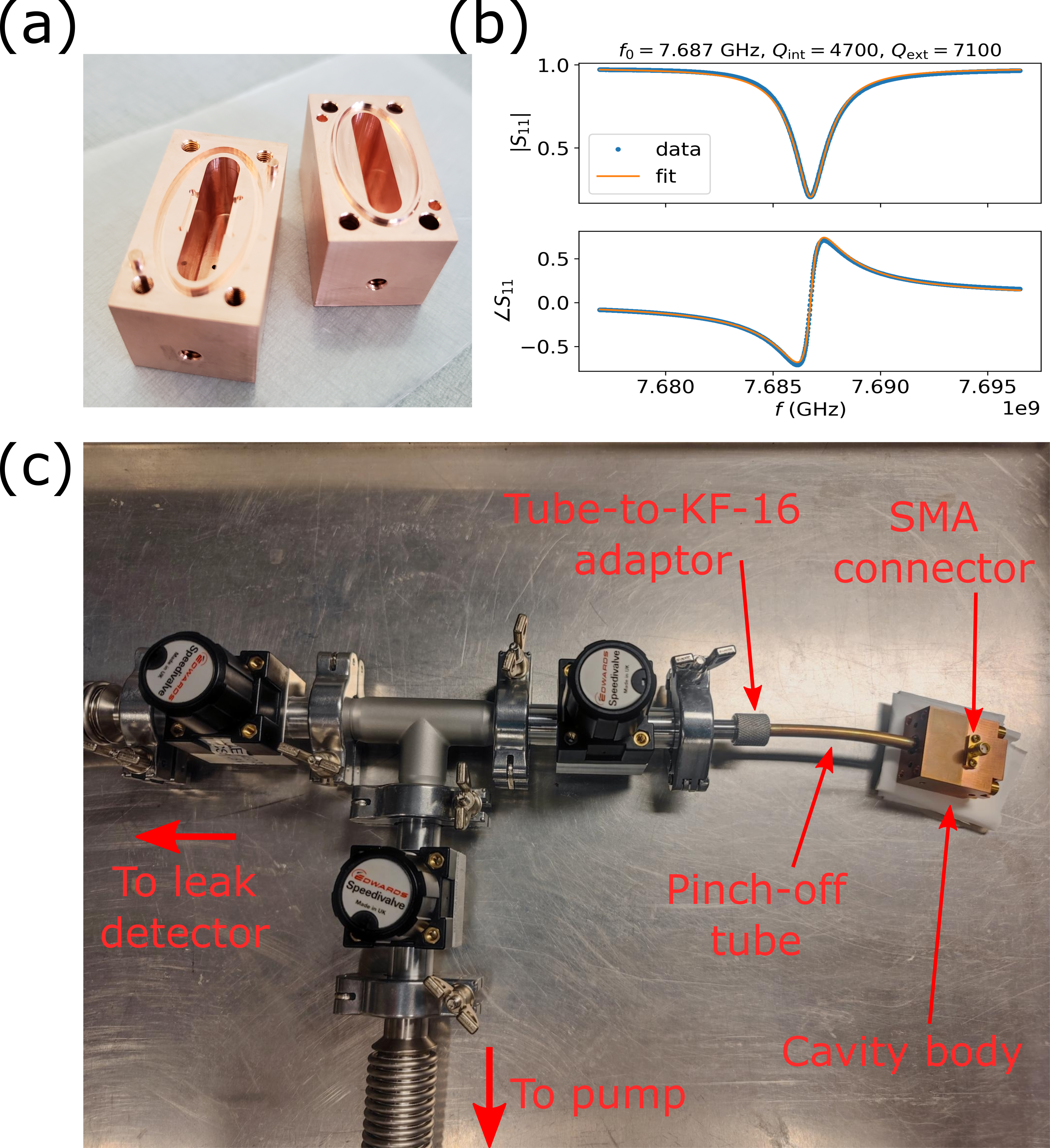}
    \caption{Fast-sealing microwave cavity. (a) Two complementary parts of the cavity body. (b) Reflection measurement of the cavity fundamental mode at room temperature. (c) Cavity vacuum connection during leak checking. The cavity connects to a vacuum assembly via its pinch-off tube and a tube-to-KF-16 adapter. Once the cavity's leak rate is validated, the pinch-off tube is crimped with a hand-held hydraulic tool, producing a seal at the pinch location.\label{fig:cavity}}
\end{figure}

For the devices, CPS resonators are chosen. These resonators consist of two parallel metal traces of width \(w\) separated by a gap \(g\) on a dielectric substrate\cite{Simons_2001} (Fig. 2). CPS resonators exhibit similar microwave properties to the more commonly tested coplanar waveguide (CPW) resonators\cite{Popovic_Nesic_1985}. Crucially, CPS resonators allow direct coupling to the cavity SMA pin, do not introduce large ground planes that distort the cavity fundamental mode, and, similar to their CPW counterparts, have participation ratios with well-understood scaling relationships with device geometry.

The design parameters of the CPS resonators are shown in Table \ref{tab:CPS}. The participation ratios per unit thickness \(\tilde{p}\) of the metal-air (MA), metal-substrate (MS), substrate-air (SA) interfaces, and the metal-air-substrate corners (C) are simulated using ANSYS Maxwell\cite{ansys}\footnotemark[1]. The devices have a fixed width of 10 \(\mathrm{\mu m}\) and a set of gaps from 10 to 100 \(\mathrm{\mu m}\). The resonator lengths are chosen so that the CPS resonator frequencies are approximately equally spaced between 4.5 and 6 GHz. A \(\lambda/4\) geometry is chosen to reduce resonator footprint. This design covers a range of participation ratios, which is useful for disentangling interface- versus non-interface-related effects\cite{Crowley_McLellan_Dutta_Shumiya_Place_Le_Gang_Madhavan_Bland_Chang_et_al._2023}.

\begin{figure}
    \centering
    \includegraphics[width=\linewidth]{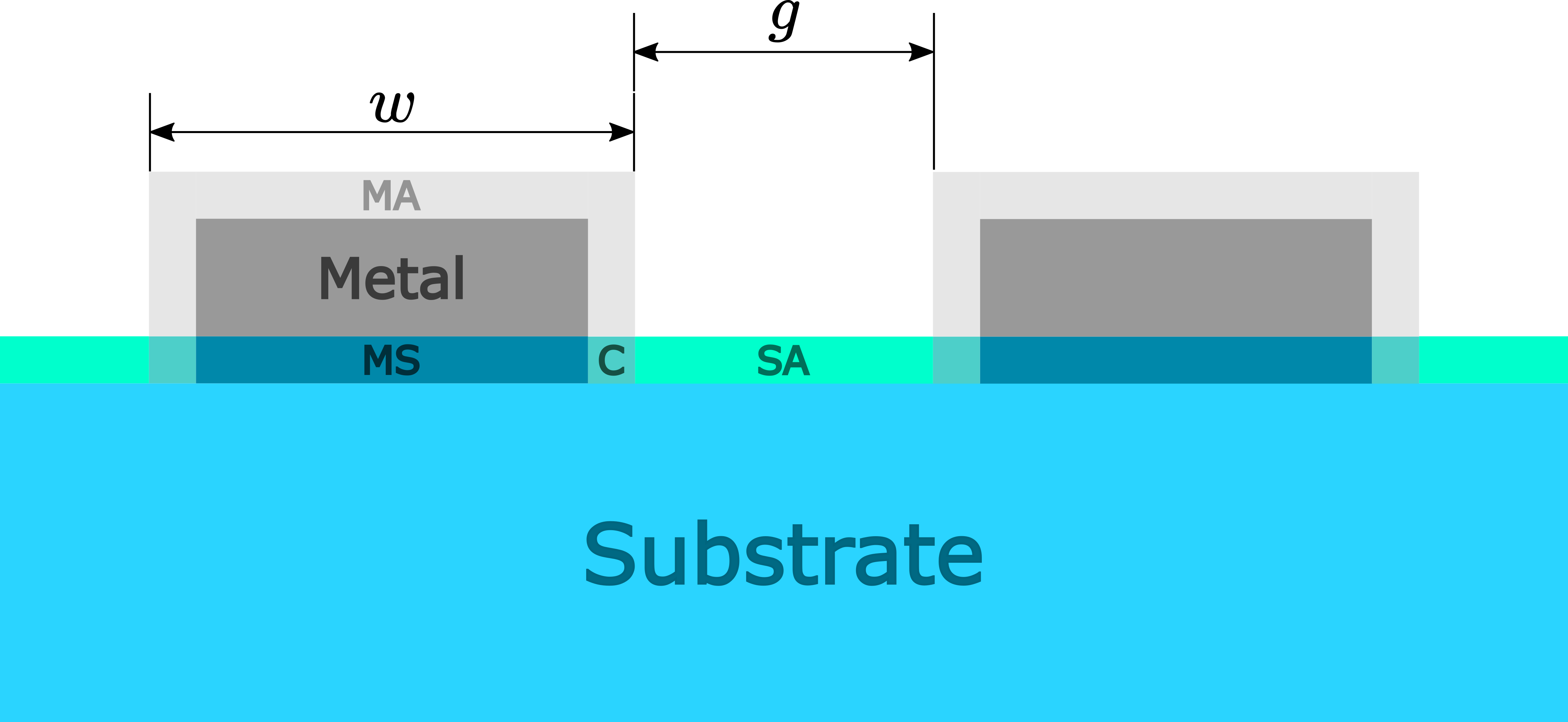}
    \caption{Cross-sectional schematic drawing of the CPS resonator. The various interface layers are labeled and exaggerated for visualization clarity.\label{fig:device}}
\end{figure}

\begin{table*}
    \centering
    \begin{ruledtabular}
        \begin{tabular}{ccccccccc}
            Index & \(w\) (\(\mathrm{\mu m}\)) & \(g\) (\(\mathrm{\mu m}\)) & Frequency (GHz) & \(\tilde{p}_\mathrm{MA}\) (ppm/nm) & \(\tilde{p}_\mathrm{MS}\) (ppm/nm) & \(\tilde{p}_\mathrm{SA}\) (ppm/nm) & \(\tilde{p}_\mathrm{c}\) (ppm/nm) & \(Q_\mathrm{ext}\) (\(\times10^6\))\\
            \colrule
            CPS1 & 10 & 10 & 4.495 & 29.7 & 196 & 192 & 41.5 & 900\\
            CPS2 & 10 & 22 & 4.986 & 21.7 & 142 & 138 & 30.4 & 330 \\
            CPS3 & 10 & 46 & 5.470 & 17.4 & 106 & 103 & 24.0 & 20\\
            CPS4 & 10 & 100 & 5.959 & 14.4 & 77.5 & 74.8 & 19.7 & 3.1\\
        \end{tabular}
    \end{ruledtabular}  
    \caption{CPS resonator design parameters from simulations.}
    \label{tab:CPS}
\end{table*}

Devices are measured in reflection mode at \(20\) mK. The fitted \(Q_\mathrm{int}\) versus the mode photon number \(n\) is plotted in Fig. 3(a). All four devices exhibit a photon-number dependence characteristic of TLS-induced loss\cite{Gao_Daal_Vayonakis_Kumar_Zmuidzinas_Sadoulet_Mazin_Day_Leduc_2008, McRae_Wang_Gao_Vissers_Brecht_Dunsworth_Pappas_Mutus_2020, Verjauw_Potocnik_Mongillo_Acharya_Mohiyaddin_Simion_Pacco_Ivanov_Wan_Vanleenhove_et_al._2021}, given by 
\begin{equation}
    \frac{1}{Q_\mathrm{int}(n, T)}=\frac{F_\mathrm{TLS}\tan\delta_\mathrm{TLS}}{\left(1+n/n_\mathrm{c}\right)^\beta}\tanh\left(\frac{\hbar\omega}{2k_\mathrm{B}T}\right)+\frac{1}{Q_\mathrm{r}} \, , \label{eq:TLS_loss}
\end{equation}
where \( F_\mathrm{TLS} \) and \( \tan\delta_\mathrm{TLS} \) are the TLS filling factor and loss tangent, \( n_\mathrm{c} \) is the critical photon number, \( \beta \) is a phenomenological parameter, and \( Q_\mathrm{r} \) is the quality factor associated with residual power-independent losses\cite{Verjauw_Potocnik_Mongillo_Acharya_Mohiyaddin_Simion_Pacco_Ivanov_Wan_Vanleenhove_et_al._2021}.

\begin{figure*}
    \centering
    \includegraphics[width=\linewidth]{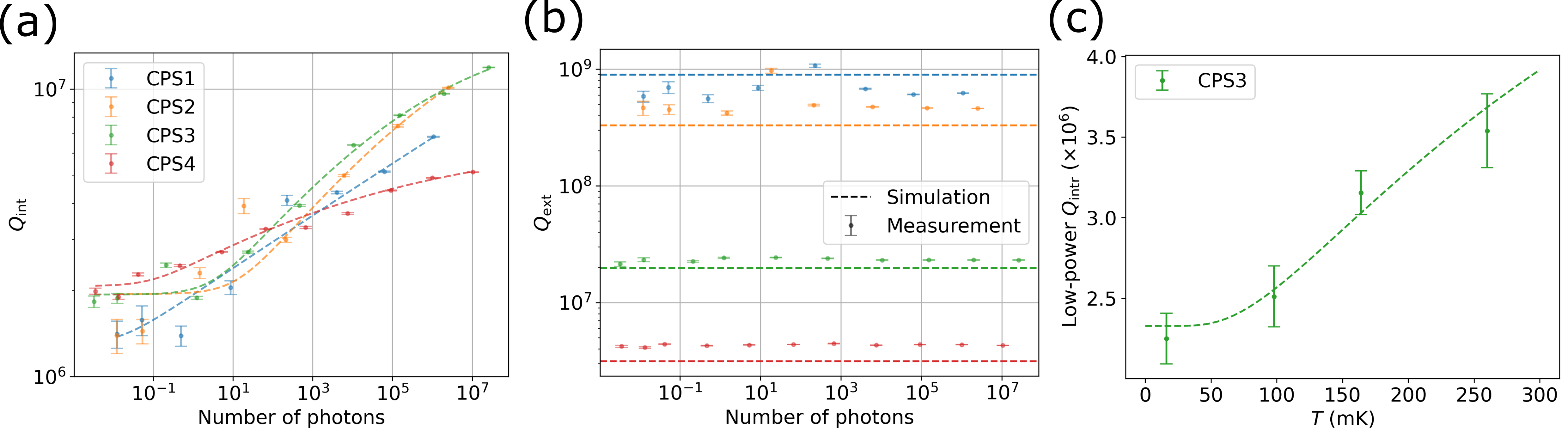}
    \caption{CPS resonator quality factor measurements. (a) CPS resonator \(Q_\mathrm{int}\) versus photon number. The lines are results of the fit to Eq. \ref{eq:TLS_loss}. (b) Simulated and measured \(Q_\mathrm{ext}\) for the CPS resonators versus resonator photon number. (c) Low-power \(Q_\mathrm{int}\) for CPS3 versus mixing chamber temperature, after corrections that remove extrinsic loss. The line is the result of the fit to Eq. \ref{eq:TLS_loss} with \(F_\mathrm{TLS}\tan\delta_\mathrm{TLS}/(1+n/n_c)^\beta\) absorbed into a single fit variable.\label{fig:Jan cooldown measurements}}
\end{figure*}

At single-photon power, all four CPS resonators exhibit \(Q_\mathrm{int}>10^6\). The low-power \(Q_\mathrm{int}\), calculated as the average \(Q_\mathrm{int}\) for \(n<n_c\), follows the expected trend for dielectric loss with the device gap dimension, where larger-footprint devices are less susceptible to dielectric loss. While the range of \(Q_\mathrm{ext}\) values spans nearly three orders of magnitude, the agreement between design and experiment values is within \(50\%\), providing a strong indication of the accuracy of numerical modeling (Fig. 3(b)).

The internal loss of the device mode is separated into two types: one type associated with the on-chip elements (e.g., dielectric loss of the various surfaces and interfaces, loss induced by quasiparticles in the superconductor), which we term the \textit{intrinsic} loss, associated with an intrinsic quality factor \(Q_\mathrm{intr}\); and the other type associated with the package (e.g., cavity wall conductor loss, Stycast dielectric loss), which we term the \textit{extrinsic} loss associated with an extrinsic quality factor \(Q_\mathrm{extr}\). Note \(1/Q_\mathrm{int}=1/Q_\mathrm{intr}+1/Q_\mathrm{extr}\).

To verify whether \(Q_\mathrm{int}\) is limited by \(Q_\mathrm{extr}\), extrinsic loss is simulated in ANSYS HFSS\footnotemark[1]. For the cavity wall conductor loss, anomalous skin effect (ASE) needs to be taken into account at \(20\) mK\cite{Chambers_1952}. Simulations show \(Q_\mathrm{extr}\) of the devices is distributed between between \(1\times 10^7\) and \(3\times 10^8\), with larger devices, closer to the cavity walls exhibiting lower \(Q_\mathrm{extr}\) (Table \ref{tab:extrinsic}). These results show \(Q_\mathrm{int}\) is dominated by \(Q_\mathrm{intr}\), and modelling of the extrinsic loss serves as a small correction except for CPS4, in which case a \(25\%\) difference between \(Q_\mathrm{int}\) and \(Q_\mathrm{intr}\) is seen. For subsequent analyses, the intrinsic quality factor \(Q_\mathrm{intr}\) is used.

\begin{table}
    \centering
    \begin{ruledtabular}
        \begin{tabular}{c>{\centering}p{2.3cm}>{\centering}p{2.3cm}>{\centering\arraybackslash}p{2.3cm}}
            Index & Measured low-power \(Q_\mathrm{int}\) (\(\times10^6\)) & Simulated \(Q_\mathrm{extr}\) (\(\times10^6\)) & Calculated low-power \(Q_\mathrm{intr}\) (\(\times10^6\))\\
            \colrule
            CPS1 & 1.5 & 250 & 1.5\\
            CPS2 & 1.7 & 340 & 1.7\\
            CPS3 & 2.0 & 97 & 2.0\\
            CPS4 & 2.0 & 12 & 2.5\\
        \end{tabular}
    \end{ruledtabular}
    
    \caption{CPS resonator \(Q_\mathrm{int}\), \(Q_\mathrm{extr}\), and \(Q_\mathrm{intr}\).}
    \label{tab:extrinsic}
\end{table}

The temperature dependence of the device quality factors is then investigated. Heat is applied to the mixing chamber (MC) of the dilution refrigerator. After the MC temperature stabilizes, an additional 30 minutes is allowed for the sample to fully thermalize before data acquisition. To minimize the time the MC is above its base operating temperature, measurements are performed on a single resonator, CPS3. The resulting temperature dependence of CPS3’s low-power \(Q_\mathrm{intr}\) is shown in Fig. 3(c). The CPS3 temperature dependence agrees well with Eq. \ref{eq:TLS_loss}, with a fit yielding \(1/F_\mathrm{TLS}\tan\delta_\mathrm{TLS}=(3.4\pm 0.6)\times 10^6\) and \(Q_\mathrm{r}=(8\pm 2)\times 10^6\). Upper bounds can be placed on each interface's thickness-loss tangent product and the film quasiparticle density, by attributing all TLS-induced loss to one interface and all power-independent loss to non-equilibrium quasiparticles, that is,
\begin{equation}
    t_i\tan\delta_i\leq \frac{F_\mathrm{TLS}\tan\delta_\mathrm{TLS}}{\tilde{p}_i} \, ,\label{eq:loss_tangent_upper_bound}
\end{equation}
where \(i\in\{\mathrm{MA}, \mathrm{MS}, \mathrm{SA}\}\)\cite{Verjauw_Potocnik_Mongillo_Acharya_Mohiyaddin_Simion_Pacco_Ivanov_Wan_Vanleenhove_et_al._2021}, and 
\begin{equation}
    n_\mathrm{qp}\leq\frac{\pi}{\alpha}\sqrt{\frac{hf_0\Delta}{2}}\frac{D\left(E_\mathrm{F}\right)}{Q_\mathrm{r}} \, ,\label{eq:qp_upper_bound}
\end{equation}
where \(n_\mathrm{qp}\) is the quasiparticle number density, \(\alpha\) is the device kinetic inductance fraction simulated using FastHenry\cite{Kamon_Ttsuk_White_1994}, \(\Delta\) is the Nb superconducting gap, and \(D(E_\mathrm{F})\) is the density of states at the Fermi surface for Nb\cite{Barends_Wenner_Lenander_Chen_Bialczak_Kelly_Lucero_O_Malley_Mariantoni_Sank_et_al._2011}. Nb material parameters for this calculation are quoted from \citet{Bonnet_Erlenkamper_Germer_Rabenhorst_1967} and \citet{Jani_Brener_Callaway_1988}.

For direct comparison with literature\cite{Verjauw_Potocnik_Mongillo_Acharya_Mohiyaddin_Simion_Pacco_Ivanov_Wan_Vanleenhove_et_al._2021, Read_Chapman_Lei_Curtis_Ganjam_Krayzman_Frunzio_Schoelkopf_2023}, the contribution from the metal-air-substrate corner is divided in half and absorbed into the MA and the MS interfaces, respectively. The estimated upper bounds are shown in Table \ref{tab:upper_bounds}. The resulting upper bounds are consistent with commonly assumed interface thicknesses\cite{Wenner_Barends_Bialczak_Chen_Kelly_Lucero_Mariantoni_Megrant_O_Malley_Sank_et_al._2011, Read_Chapman_Lei_Curtis_Ganjam_Krayzman_Frunzio_Schoelkopf_2023} and previously reported interface loss tangents\cite{Verjauw_Potocnik_Mongillo_Acharya_Mohiyaddin_Simion_Pacco_Ivanov_Wan_Vanleenhove_et_al._2021, Read_Chapman_Lei_Curtis_Ganjam_Krayzman_Frunzio_Schoelkopf_2023}, and are comparable to previously reported non-equilibrium quasiparticle density in Nb films\cite{Noguchi_Mima_Otani_2021}.

\begin{table}
    \centering
    \begin{ruledtabular}
        \begin{tabular}{p{3.4cm}>{\centering}p{1.5cm}>{\centering}p{1.5cm}>{\centering\arraybackslash}p{1.5cm}}
             & MA (including half corner) & MS (including half corner) & SA\\
            \colrule
            \multicolumn{4}{l}{\textit{Interface losses}}\\
            Upper bound on \(t_i\tan\delta_i\) (\(\times 10^{-3}\cdot\mathrm{nm}\)) & \(10 \pm 2\) & \(2.5\pm 0.5\) & \(2.9\pm 0.5\)\\
            Literature \(\tan\delta_i\) (\(\times 10^{-3}\)) & \(9.9\) & \textemdash & \(< 1.1\)\\
            Upper bound on \(t_i\) (nm) & \(1.0 \pm 0.2\) & \textemdash & \(2.6 \pm 0.5\)\\
            \colrule
            \multicolumn{4}{l}{\textit{Quasiparticles}}\\
            Upper bound on \(n_{\mathrm{qp}}\) (\(\mathrm{\mu m}^{-3}\)) & \multicolumn{3}{c}{\(400 \pm 100\)} \\
            Literature \(n_{\mathrm{qp}}\) (\(\mathrm{\mu m}^{-3}\)) & \multicolumn{3}{c}{\(\sim 100\)}\\
        \end{tabular}
    \end{ruledtabular}
    \caption{Upper bounds on the interface thickness-loss tangent product and the quasiparticle density. "Half corner" indicates that half of the metal–air–substrate corner contribution is absorbed into MA and MS, respectively. The literature values for \(\tan\delta_\mathrm{MA}\), \(\tan\delta_\mathrm{SA}\), and \(n_\mathrm{qp}\) are taken from \citet{Verjauw_Potocnik_Mongillo_Acharya_Mohiyaddin_Simion_Pacco_Ivanov_Wan_Vanleenhove_et_al._2021}, \citet{Read_Chapman_Lei_Curtis_Ganjam_Krayzman_Frunzio_Schoelkopf_2023}, and \citet{Noguchi_Mima_Otani_2021}, respectively.}
    \label{tab:upper_bounds}
\end{table}

Although there are four measurements from four devices and four unknowns (\(t_\mathrm{MA}\tan\delta_\mathrm{MA}\), \(t_\mathrm{MS}\tan\delta_\mathrm{MS}\), \(t_\mathrm{SA}\tan\delta_\mathrm{SA}\), \(n_\mathrm{qp}\)), a direct inversion of the system of equations \(1/Q_{\mathrm{intr},j}=\sum_i\tilde{p}_{i,j}t_{i,j}\tan\delta_{i,j}+(\alpha_j/\pi D(E_\mathrm{F}))\sqrt{2/hf_0\Delta}n_\mathrm{qp}\), where \(j\) is the device index, does not yield reliable estimates of these unknown variables. The limitation is that the participation ratios of different interfaces vary nearly collinearly across different devices, so their respective contributions to loss cannot be well separated. A better-conditioned inversion would require a device set with more diverse weights in the participation ratios of different interfaces\cite{Woods_Calusine_Melville_Sevi_Golden_Kim_Rosenberg_Yoder_Oliver_2019}. We defer a full separation of each individual interface loss to future work.

To isolate the role of the MA interface, the sample is warmed to room temperature, and the pinch-off tube is cut open, exposing the sample to ambient conditions for 8 days. The devices are then cooled down and re-measured. Across all four resonators, \(Q_\mathrm{intr}\) decreased by around \(30\%\) (Fig. 4(a)), while \(Q_\mathrm{ext}\) is largely unchanged. Systematic trends in the shifts of device \(f_0\) and \(Q_\mathrm{intr}\) with device gap dimensions are observed. These shifts can be understood as consequences of oxide regrowth at the MA interface. Regrown Nb oxides increase the amount of dielectric material between the metal traces and decrease the height of the metal traces, thus increasing the resonator capacitance and inductance and lowering \(f_0\). \(Q_\mathrm{intr}\) degrades as more lossy materials are formed. The \(Q_\mathrm{intr}\) degradation observed after air re-exposure underscores the effectiveness of the initial HF acid etching and sample packaging with the fast-sealing microwave cavity.

\begin{figure*}
    \centering
    \includegraphics[width=\linewidth]{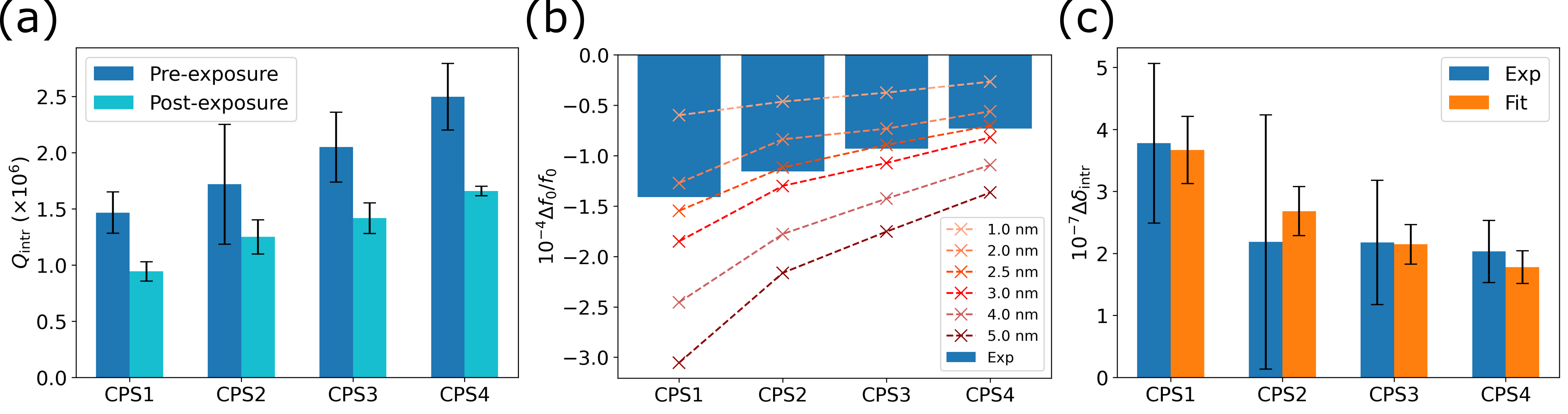}
    \caption{CPS resonator quality factor measurements before and after re-oxidation. (a) CPS resonator intrinsic quality factor \(Q_\mathrm{intr}\) before and after 8 days of air re-exposure. (b) Measured resonance frequency shifts of CPS resonators, along with simulated shifts for different assumed oxide thicknesses.(c) Measured and fitted changes in \(\delta_\mathrm{intr}\) with \(\Delta t_\mathrm{MA}\) set to 2.5 nm. Extracted \(\tan\delta_\mathrm{MA}=(2.9\pm0.7)\times 10^{-3}\). \label{fig:Feb cooldown measurements}}
\end{figure*}

To understand the oxide regrowth quantitatively, simulations are performed\footnotemark[1] with a simplifying assumption that only Nb\(_2\)O\(_5\) forms between the two cooldowns. This assumption lets the increase in oxide thickness and the concomitant decrease in Nb thickness follow the stoichiometric relationship. ANSYS Maxwell is used to simulate the capacitance change versus oxide regrowth thickness, and FastHenry\cite{Kamon_Ttsuk_White_1994} is used to model the inductance change from the corresponding reduction in Nb thickness. Combining these effects yields the predicted frequency shifts at different oxide regrowth thicknesses (Fig. 4(b)). A regrowth thickness \(\Delta t_\mathrm{MA}=2.5\) nm best matches the experimental data. This result is comparable to the \(1.8\) nm predicted for 8 days of exposure, from the Nb\(_2\)O\(_5\) growth kinetics measured by \citet{Verjauw_Potocnik_Mongillo_Acharya_Mohiyaddin_Simion_Pacco_Ivanov_Wan_Vanleenhove_et_al._2021}.

With \(\Delta t_\mathrm{MA}\) determined, by attributing the additional loss seen in the second cooldown solely to the regrown Nb oxide, the loss tangent of the MA interface can be extracted via \(\tan\delta_\mathrm{MA}=\Delta\delta_\mathrm{intr}/\left(\tilde{p}_\mathrm{MA}\Delta t_\mathrm{MA}\right)\), where \(\Delta\delta_\mathrm{intr}\) is the difference in \(1/Q_\mathrm{intr}\) between the two cooldowns (Fig. 4(c)). Averaging over all four resonators gives \(\tan\delta_\mathrm{MA}=(2.9\pm0.7)\times10^{-3}\). The averaged loss tangent returns \(\Delta\delta_\mathrm{intr}\) consistent with measurements within error bars for all four devices. This result is comparable to \(\tan\delta_\mathrm{MA}=9.9\times10^{-3}\) reported by \citet{Verjauw_Potocnik_Mongillo_Acharya_Mohiyaddin_Simion_Pacco_Ivanov_Wan_Vanleenhove_et_al._2021}, and consistent with the previously determined thickness-loss tangent product upper bound in Table \ref{tab:upper_bounds}.

In summary, we have developed a fast-sealing microwave cavity that provides a practical and consistent solution to the time-critical workflow for superconducting device experiments where suppression of oxide regrowth is crucial. Samples can be loaded and evacuated consistently within 5 minutes after the final oxide etch, enabling reproducible measurements and facilitating controlled studies of surface re-oxidation. By deliberately re-oxidizing the sample in air, we connected measured device resonance frequency shifts quantitatively to Nb\(_2\)O\(_5\) regrowth, which, when combined with the corresponding degradation in device \(Q_\mathrm{intr}\), allowed the extraction of the MA interface loss tangent. While the present devices are limited by having similar weights in interface participation ratios, and cannot be used to reliably disentangle loss from individual interfaces, upper bounds on different interface thickness-loss tangent products and the non-equilibrium quasiparticle density are consistent with literature values\cite{Verjauw_Potocnik_Mongillo_Acharya_Mohiyaddin_Simion_Pacco_Ivanov_Wan_Vanleenhove_et_al._2021, Read_Chapman_Lei_Curtis_Ganjam_Krayzman_Frunzio_Schoelkopf_2023, Noguchi_Mima_Otani_2021}. The fast-sealing microwave cavity opens a new pathway to study interface-related losses, the understanding of which is critical in pushing for higher-coherence superconducting quantum devices.

\begin{acknowledgments}
We would like to thank Saba Sadeghi, Nina Heinig, Joseph Thomas, and George Nichols for assisting with film growth and characterization, the University of Waterloo Quantum Nanofab
team members for assistance on the device fabrication, and Anna Grassellino for helpful discussions.

This material is based upon work supported by the U.S. Department of Energy, Office of Science, under Award Number FNAL 23-02 and the U.S. Department of Energy, Office of Science, Office of Nuclear Physics under contract DE-AC05-06OR23177. M.B. also acknowledges support from U.S. Department of Energy, Office of Science, National Quantum Information Science Research Centers, Superconducting Quantum Materials and Systems Center (SQMS), under Contract No. 89243024CSC000002. Fermilab is operated by Fermi Forward Discovery Group, LLC under Contract No. 89243024CSC000002 with the U.S. Department of Energy, Office of Science, Office of High Energy Physics.

We also acknowledge support from the Natural Sciences and Engineering Research Council (NSERC) and the Canadian Microelectronics Corporation (CMC).
\end{acknowledgments}

\bibliography{references}

\end{document}


\preprint{AIP/123-QED}

\title[Supplementary information for "Investigation of coherence of niobium-based resonators enabled by a fast-sealing microwave cavity"]{Supplementary information for "Investigation of coherence of niobium-based resonators enabled by a fast-sealing microwave cavity"}
\author{C. Zhang}
\author{R. Germond}%
\author{N. Janzen}
\affiliation{Institute for Quantum Computing, University of Waterloo, Waterloo, ON N2L 3G1, Canada}%
\affiliation{Department of Physics and Astronomy, University of Waterloo, ON N2L 3G1, Canada}

\author{M. Bal}
\affiliation{Superconducting Quantum Materials and Systems Division, Fermi National Accelerator Laboratory (FNAL), Batavia, IL 60510, USA.}%

\author{A.-M. Valente-Feliciano}
\affiliation{Jefferson Lab, 12000 Jefferson Ave, Newport News, VA 23606, USA.}

\author{A. Lupascu}
  \email{c76zhang@uwaterloo.ca, adrian.lupascu@uwaterloo.ca}
\affiliation{Institute for Quantum Computing, University of Waterloo, Waterloo, ON N2L 3G1, Canada}%
\affiliation{Department of Physics and Astronomy, University of Waterloo, ON N2L 3G1, Canada}
\affiliation{Waterloo Institute for Nanotechnology, University of Waterloo, Waterloo, ON, N2L 3G1, Canada}

\date{\today}

\maketitle

\section{Sample fabrication}
Resonators are fabricated on a 2'' c-plane HEMEX-grade sapphire wafer, selected for its low loss tangent\cite{Read_Chapman_Lei_Curtis_Ganjam_Krayzman_Frunzio_Schoelkopf_2023}. The wafer substrate is first sonicated in acetone, isopropyl alcohol (IPA), and deionized (DI) water, then loaded into a custom sputtering chamber, pumped overnight as the wafer is heated to 800 \(^\circ\)C and let cool down to reach a pressure \(<10^{-10}\) mbar. Niobium (Nb) is sputtered at 770 \(^\circ\)C to a nominal thickness of 150 nm (145 nm measured). The wafer is then vapor-primed with hexamethyldisilazane, spin-coated with Microposit S1811 photoresist, baked at 120 \(^\circ\)C, exposed with a 405-nm laser beam, and developed in Microposit MF-319. Next, the wafer is ashed with oxygen plasma, followed by reactive-ion etching (RIE) in a 3:1:1 Cl\(_2\):BCl\(_3\):Ar gas mixture. The etching time is set to be \(25\%\) longer than required for etching 145 nm of Nb to ensure its complete removal. The etched wafer is immediately rinsed in DI water for 1 min to remove reactive RIE byproducts and suppress film corrosion\cite{Anferov_Lee_Zhao_Simon_Schuster_2024}. Resist is stripped by soaking overnight in Remover PG, followed by sonication in fresh Remover PG at 80 \(^\circ\)C, acetone and IPA at room temperature, and oxygen plasma ashing. After optical inspection, AZ P4620 photoresist is spin-coated as a protective layer, and the wafer is diced into \(7.5~\mathrm{mm}\times 7.5~\mathrm{mm}\) dies. A die selected for experiments is resist-stripped, dipped in a 1:10 HF:DI water solution for 1 min, rinsed in DI water baths, blow-dried, and then loaded and sealed in the fast-sealing microwave cavity.

\section{Thin-film characterization}
\begin{figure}
    \centering
    \includegraphics[width=\linewidth]{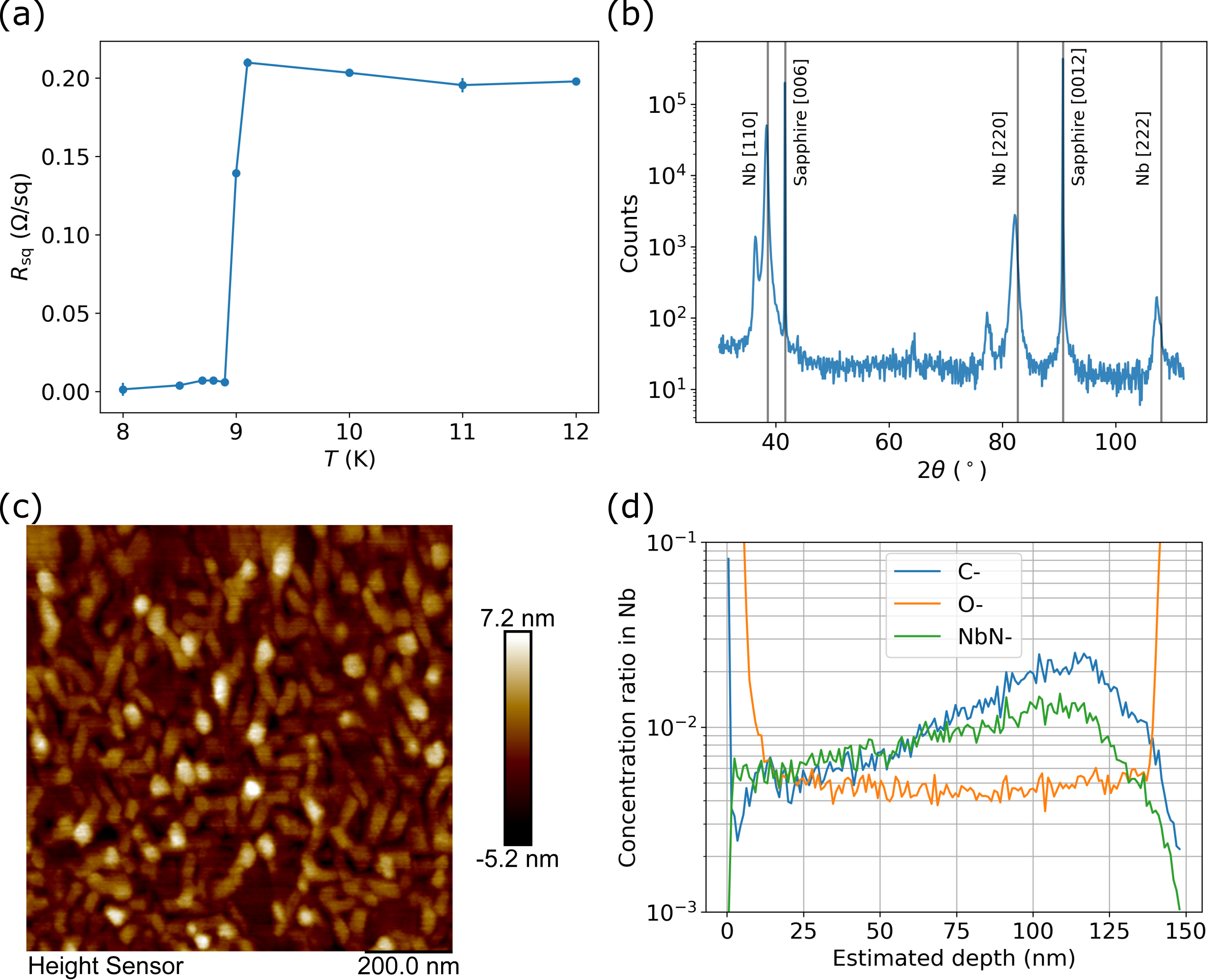}
    \caption{Characterization of the Nb thin film. (a) Nb sheet resistance \(R_\mathrm{sq}\) versus cryostat temperature \(T\). The critical temperature is \(T_c=9.0\) K. (b) \(2\theta\)-\(\omega\) scan of the device film, indicating a polycrystalline film with predominantly (110) texture. (c) Atomic force microscopy (AFM) topograph of the Nb surface. The elongated-grain morphology is consistent with the (110) Nb texture observed by XRD. (d) Time-of-flight secondary ion mass spectrometry (ToF-SIMS) depth profiling analysis of the device film. A \(\sim70\) nm region is observed where the carbon atomic concentration is \(>1\%\).\label{fig:characterization}}
\end{figure}

Thin films from the same wafer as the measured devices are characterized with low-temperature DC measurements, X-ray diffraction (XRD), atomic force microscopy (AFM), and time-of-flight secondary ion mass spectrometry (ToF-SIMS) (Fig. \ref{fig:characterization}).

DC measurements in a van der Pauw configuration\cite{Van_Der_Pauw_1991} show a critical temperature \(T_c=9.0\) K and a residual resistivity ratio \(\mathrm{RRR}=7.2\).

XRD indicates a polycrystalline film with predominantly [110] orientation. Nb films grown on c-plane sapphire under similar conditions have been reported to adopt either [110]\cite{Cuomo_Angilello_1973, Dietrich_Boyen_Koslowski_2003} or [111]\cite{Drimmer_Telkamp_Fischer_Rodrigues_Todt_Krizek_Kriegner_Muller_Wegscheider_Chu_2024, McFadden_Oh_Zhou_Larson_Gill_Dixit_Simmonds_Lecocq_2024} orientations. In particular, in \citet{Cuomo_Angilello_1973}, [110] growth was observed on the substrate with few dislocations, and [111] growth was observed on the substrate with higher dislocation density.

AFM shows an elongated-grain structure characteristic of [110] growth\cite{Drimmer_Telkamp_Fischer_Rodrigues_Todt_Krizek_Kriegner_Muller_Wegscheider_Chu_2024, Premkumar_Weiland_Hwang_Jack_Place_Waluyo_Hunt_Bisogni_Pelliciari_Barbour_et_al._2021}, consistent with the XRD measurement. The root-mean-square (\(R_q\)) and arithmetic-mean (\(R_a\)) roughness values are \(1.72\) nm and \(1.35\) nm.

For ToF-SIMS, depth profiling is performed. Using relative sensitivity factors (RSFs) from the literature\cite{Maheshwari_2012, Angle_Lechner_Palczewski_Reece_Stevie_Kelley_2022, Maheshwari_Stevie_Myeneni_Ciovati_Rigsbee_Griffis_2011}, various impurity concentrations can be estimated\cite{Vickerman_Gilmore_2009}. The measurement reveals a \(\sim\!70~\mathrm{nm}\) region near the metal–substrate (MS) interface containing \(>1\%\) atomic concentration of carbon. While this concentration is higher than anticipated, it is not expected to impact the oxidation dynamics relevant for the scope of this work.

\section{Resonator microwave characterization setup}

\begin{figure}
    \centering
    \includegraphics[width=\linewidth]{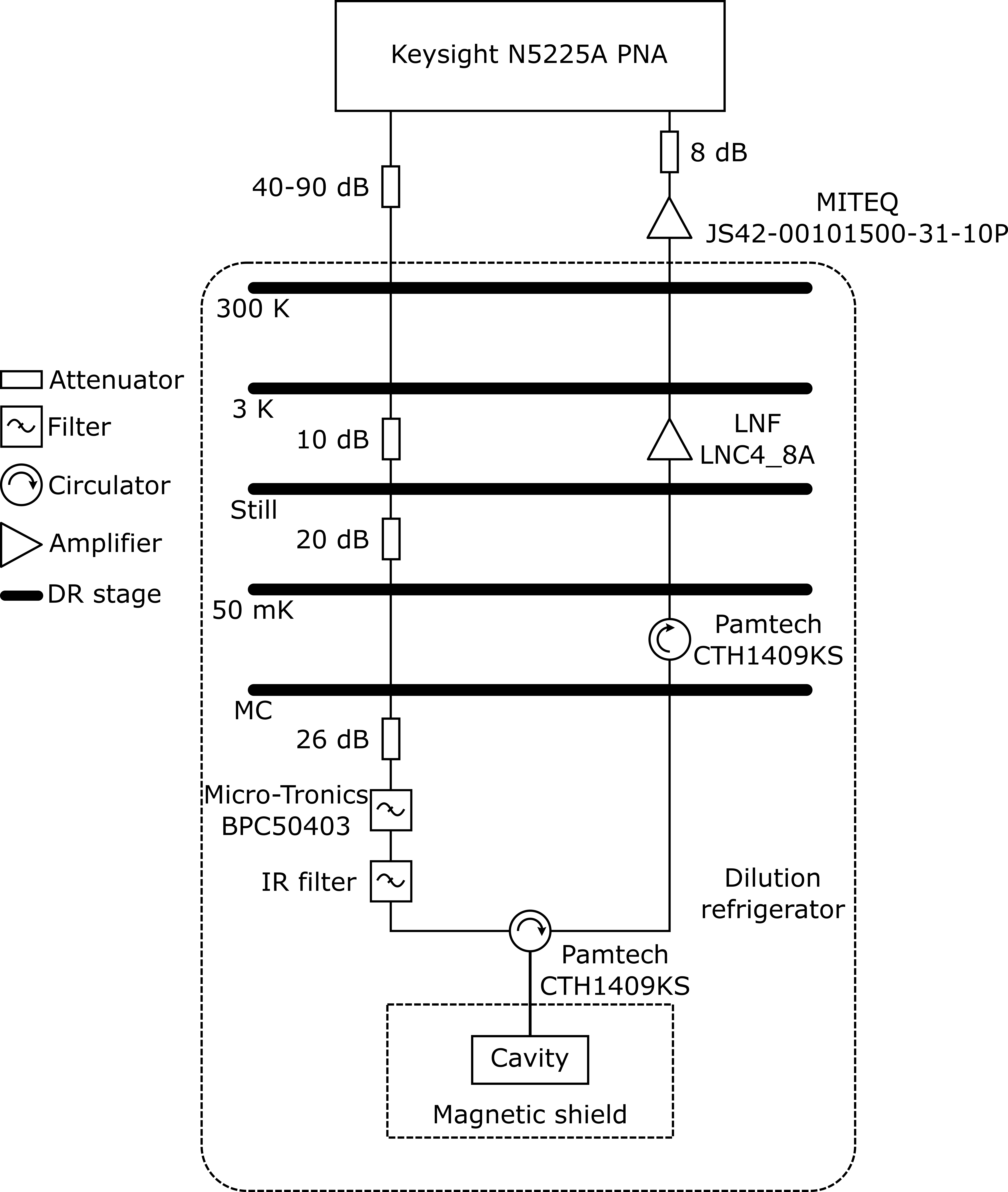}
    \caption{Schematic wiring diagram for resonator reflection measurement.\label{fig:setup}}
\end{figure}

Fig. \ref{fig:setup} shows the chain of microwave components used to measure the reflection signal of the coplanar stripline (CPS) resonators. All filters, circulators, and amplifiers are specified to operate in the 4-8 GHz band or wider. A Keysight N5225A PNA vector network analyzer is used. The choice of room-temperature attenuation (ranging from 40 to 90 dB), together with the VNA source power (ranging from -30 to 20 dBm), sets a ten-order-of-magnitude span of input power at the cavity. The reflected signal is routed via a Pamtech cryogenic circulator to the amplifiers before returning to the VNA receiver. Power at the device plane is estimated from known insertion losses and gains from the component datasheets or our own previous measurements.

\section{\texorpdfstring{ANSYS HFSS simulations: resonator mode \(Q_\mathrm{ext}\) and \(Q_\mathrm{extr}\)}{ANSYS HFSS simulations: resonator mode Qext and Qextr}}

\begin{figure}
    \centering
    \includegraphics[width=\linewidth]{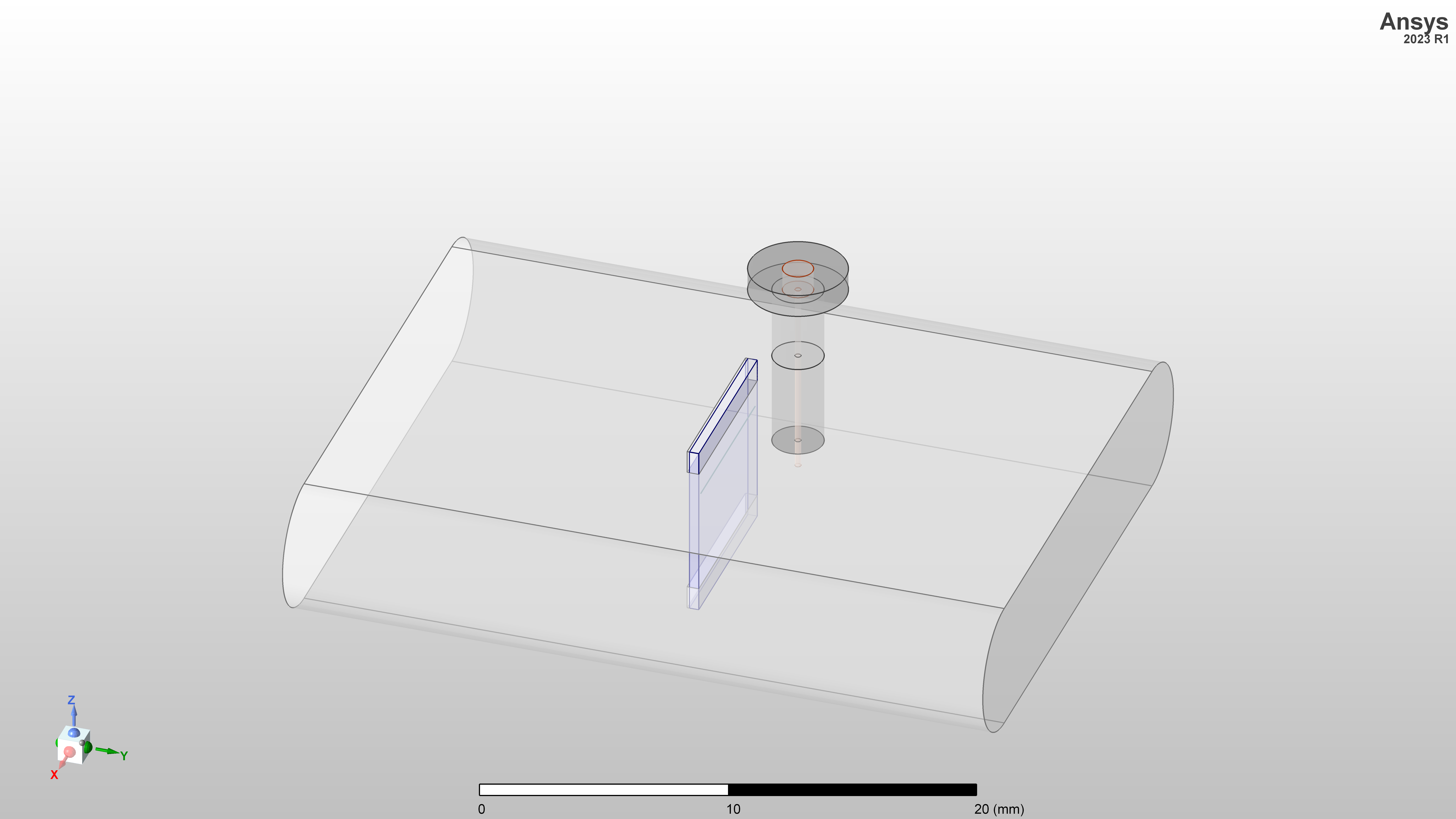}
    \caption{ANSYS HFSS simulation setup for \(Q_\mathrm{ext}\) and \(Q_\mathrm{extr}\) simulations. The device on sapphire, the cavity inner volume, and the SMA connector port, filled with Stycast 2850FT, are modeled explicitly.\label{fig:HFSS_sim}}
\end{figure}

To establish expectations for the resonator mode linewidths and suppression levels, \(Q_\mathrm{ext}\) needs to be simulated for the devices at their nominal positions. To correct for extrinsic losses arising from off-chip elements (e.g., cavity wall conductor loss and Stycast dielectric loss), \(Q_\mathrm{extr}\) needs to be simulated. Both quantities can be extracted from a single HFSS\cite{ansys} model where all on-chip elements (e.g., substrate, Nb, and material interfaces) are treated as lossless (Fig. \ref{fig:HFSS_sim}).

However, these simulations prove difficult to converge due to (i) large separation of scales ((\(\mu\)m-sized resonators versus cm-sized cavity internal space), (ii) extremely narrow linewidths of the simulated modes (down to \(20\) Hz), and (iii) drifts in simulated resonance frequency with increasing mesh density. To be able to locate and resolve the resonance dip, the following two-stage simulation workflow is designed:
\begin{enumerate}
    \item Stage I: an eigenmode simulation is performed to refine the mesh at the device traces and the cavity port, and to obtain an initial resonance frequency estimate.
    \item Stage II: a modal network simulation is performed over the expected dip, with its initial mesh imported from stage I.
\end{enumerate}
All four device modes are located and resolved with this approach.

At low temperature, copper is also known to exhibit anomalous skin effect (ASE) as its electron mean free path exceeds its classically predicted skin depth\cite{Reuter_Sondheimer_1948, Chambers_1950, Chambers_1952}. In this regime, the surface resistance is found to scale as \(R\propto \omega^{2/3}\), rather than \(R\propto \omega^{1/2}\) as predicted by the classical skin-effect theory, where \(\omega\) is the measurement frequency. It is convenient to infer the surface resistance of the cavity walls at a resonator mode frequency based on an empirical \(R_\mathrm{meas, c}=4.60\times10^{-3}\) \(\Omega/\square\), extracted from the cavity-mode \(Q_\mathrm{int}\) measured at the base temperature. The frequency scaling relationship is given by
\begin{equation}
    R_{\mathrm{cl}, r}=R_{\mathrm{meas}, c}\left(\frac{\omega_r}{\omega_c}\right)^\frac{1}{2} \, ,
\end{equation}
from the classical skin effect, and
\begin{equation}
    R_{\mathrm{ASE}, r}=R_{\mathrm{meas}, c}\left(\frac{\omega_r}{\omega_c}\right)^\frac{2}{3} \, ,
\end{equation}
from the anomalous skin effect, where \(R_{\mathrm{cl}, r}\) and \(R_{\mathrm{ASE}, r}\) are the cavity wall surface resistance at a resonator mode, calculated from the classical and anomalous skin effect, respectively, and \(\omega_r\) and \(\omega_c\) are the resonator and cavity mode frequencies.

In HFSS, the "finite conductivity" boundary condition is used to capture the cavity wall conductor loss. This model treats the skin effect classically. Thus, an additional correction from the ASE is added by hand,
\begin{equation}
    R_{\mathrm{ASE}, r}=R_{\mathrm{cl}, r}\left(\frac{\omega_r}{\omega_c}\right)^\frac{1}{6} \, ,
\end{equation}
and as the quality factor associated with wall conductor loss, which dominates \(Q_\mathrm{extr}\), scales\cite{Pozar_2012} as \(1/R\), it follows that
\begin{equation}
    Q_\mathrm{extr, ASE} = Q_\mathrm{extr, cl} \left(\frac{\omega_c}{\omega_r}\right)^\frac{1}{6} \, ,
\end{equation}
where \(Q_\mathrm{extr, cl}\) is the extrinsic quality factor extracted from HFSS with classical skin effect, and \(Q_\mathrm{extr, ASE}\) is the ASE-corrected extrinsic quality factor.

\section{ANSYS Maxwell simulations: interface participation ratios and resonator capacitance shifts from oxide regrowth}

\begin{figure}
    \centering
    \includegraphics[width=\linewidth]{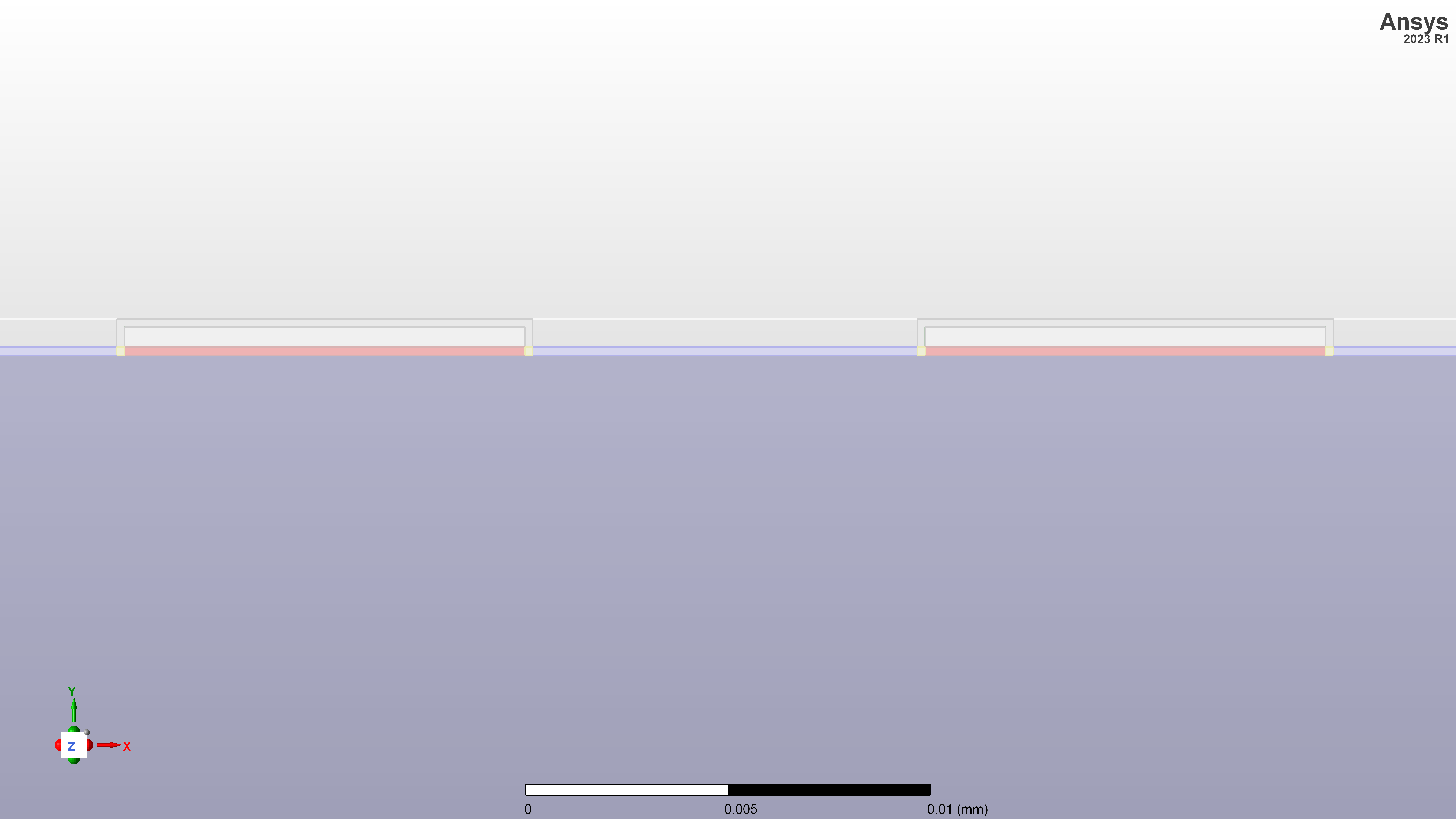}
    \caption{ANSYS Maxwell 2D simulation setup. Interface thicknesses are exaggerated for visualization. The metal traces are shown in white, and the substrate is in dark blue. The metal-air (MA), metal-substrate (MS), substrate-air (SA), and metal-air-substrate corners (C) are colored gray, red, blue, and yellow, respectively. Interface layers are assigned a dielectric constant \(\epsilon_r=10\), as per \citet{Read_Chapman_Lei_Curtis_Ganjam_Krayzman_Frunzio_Schoelkopf_2023}. Equal and opposite voltages are applied to the two metal traces.\label{fig:Maxwell_sims}}
\end{figure}

Interface participation ratios and oxide regrowth-induced resonator capacitance shifts are obtained from 2D electrostatic simulations in ANSYS Maxwell\cite{ansys} (Fig. \ref{fig:Maxwell_sims}).

Because the goal is to capture relative frequency shifts on the order of \(10^{-4}\), the simulated relative capacitance changes need to be precise to \(\sim 10^{-5}\). Convergence to this level requires a large portion of the substrate (\(\sim 100\) \(\mu\)m to \(\sim 1\) mm) to be included, making the nm-scale interface layers too thin to be meshed. We validate and rely on the linear proportionality of both the interface participation ratios and the relative capcitance changes with interface thickness \(t_i\) (for \(t_i\ll t_\mathrm{Nb}\), where \(t_\mathrm{Nb}\) is the metal thickness). The following simulation procedure is then used.

\begin{itemize}
    \item \textbf{Interface participation ratios:}
    \begin{enumerate}
        \item Choose the smallest interface thickness that can be meshed by Maxwell.
        \item Compute the participation ratio \(p_i\) of interface \(i\in\{\mathrm{MA}, \mathrm{MS}, \mathrm{SA}, \mathrm{C}\}\) in the ANSYS built-in Fields Calculator, where \(p_i=\int_{S_i}ds\, \epsilon_r|\mathbf{E}|^2/\int_{S_\mathrm{tot}}ds\, \epsilon_r|\mathbf{E}|^2\), \(\epsilon_r\) is the relative permittivity, \(\mathbf{E}\) is the simulated electric field, and \(S_i\) and \(S_\mathrm{tot}\) are regions representing the interface \(i\) and the full simulation, respectively.
        \item Calculate the participation ratio per unit thickness as \(\tilde{p}_i=p_i/t_i\), where \(t_i\) is the thickness of the interface \(i\) in the simulation.
    \end{enumerate}
    \item \textbf{Relative capacitance shifts:}
    \begin{enumerate}
        \item Simulate the baseline capacitance per unit length for \(t_\mathrm{Nb}=145\) nm and \(t_\mathrm{MA}=0\), using \(C=2W_e^2/V^2\) and \(W_e=\int_{S_\mathrm{tot}}ds\, \epsilon_r|\mathbf{E}|^2/2\), where \(V\) is the potential difference between the two CPS metal traces.
        \item For each oxide regrowth thickness \(\Delta t_\mathrm{MA}\), calculate the corresponding Nb consumption with \(\Delta t_\mathrm{Nb}=\beta\rho_\mathrm{Nb_2O_5}\Delta t_\mathrm{MA}/\rho_\mathrm{Nb}\) and \(\beta=2A_\mathrm{Nb}/\left(2A_\mathrm{Nb}+5A_\mathrm{O}\right)\), where \(A\) is the mass number of the element, quoted from \citet{De_Laeter_Bohlke_De_Bievre_Hidaka_Peiser_Rosman_Taylor_2003}, and \(\rho\) is the density of the substance, quoted from \citet{Haynes_2014}.
        \item Simulate the capacitance per unit length \(C^\prime\) with \(t_\mathrm{Nb}^\prime=t_\mathrm{Nb}-\Delta t_\mathrm{Nb}\) and \(t_\mathrm{MA}=0\).
        \item Simulate the capacitance per unit length \(C^{\prime\prime}\) with  \(t_\mathrm{Nb}^\prime=t_\mathrm{Nb}-\Delta t_\mathrm{Nb}\) and \(t_\mathrm{MA}=t_\mathrm{min}\), where \(t_\mathrm{min}\) is the thinnest oxide thickness that can be meshed by ANSYS Maxwell. Define \(\gamma=(C^{\prime\prime}-C^\prime)/C^\prime\).
        \item Using linear proportionality of \(\gamma\) with \(t_\mathrm{MA}\), calculate the relative capacitance shift from the specified oxide regrowth thickness \(\gamma^\prime=\Delta t_\mathrm{MA}\gamma/t_\mathrm{min}\).
        \item Calculate the capacitance per unit length with \(\Delta t_\mathrm{Nb}\) decrease in metal thickness and \(\Delta t_\mathrm{MA}\) oxide regrowth as \(C^{\prime\prime\prime}=\left(1+\gamma^\prime\right)C^\prime\).
        \item Calculate capacitance shift for a given \(\Delta t_\mathrm{MA}\) as \(\left(C^{\prime\prime\prime}-C\right)/C\).
    \end{enumerate}
\end{itemize}

\section{FastHenry simulations: CPS resonator kinetic inductance fraction and inductance shifts from oxide regrowth}

For inductance simulations, FastHenry\cite{Kamon_Ttsuk_White_1994} is used because it can model both the kinetic inductance of superconducting materials and the finite thickness of metal traces. FastHenry does not include the dielectric substrate, but as the magnetic susceptibility of sapphire is very small \cite{Gupta_2007} (\(\chi_m=-14\times 10^{-6}\)), the absence of the substrate does not significantly impact the results of inductance simulations.

Striplines of the nominal device length \(l\), width \(w\), and gap \(g\) are modeled. The striplines are modeled to be open-ended on both ends. For each stripline, a differential voltage excitation is applied between its two end terminals. FastHenry returns the impedance matrix \(Z\) of the setup. Our simulation setups are lossless, and run at the resonant frequencies \(\omega_0\) of the resonators. Thus, the inductance matrix can be calculated from \(Z\), given by:
\begin{equation}
    L_\mathrm{mat}=\frac{Z}{j\omega_0}=\begin{pmatrix}
L_1 & M\\
M & L_2
\end{pmatrix} \, ,
\end{equation}
where \(L_1\) and \(L_2\) are the self-inductances of the individual metal striplines, and \(M\) the mutual inductance between the two metal strips. The effective inductance per unit length is then \(L=L_\mathrm{eff}/l\), where \(L_\mathrm{eff}=L_1+L_2-2M\).

The kinetic inductance fraction is defined as the ratio of kinetic inductance to total inductance of a device\cite{Barends_2009} \(\alpha=L_k/L_\mathrm{tot}\). \(\alpha\) is simulated by comparing the inductance \(L_{(1)}\) of normal and nearly lossless metal with conductivity \(\sigma=1\times 10^{30}\) S/m, versus the inductance \(L_{(2)}\) of a superconductor with London penetration depth \(\lambda=39\) nm\cite{Maxfield_McLean_1965}. Combining the two results gives \(\alpha=\left(L_{(1)}-L_{(2)}\right)/L_{(1)}\).

The inductance for each oxide regrowth thickness is simulated using the metal thickness \(t_\mathrm{Nb}^\prime=t_\mathrm{Nb}-\Delta t_\mathrm{Nb}\), calculated as detailed in Section IV.

\bibliography{references}